\documentclass[10pt,twocolumns,article,nofontune,letter]{IEEEtran}

\usepackage{url}
\usepackage[utf8]{inputenc}
\usepackage{xcolor}
\usepackage{amsmath}
\usepackage{amssymb}
\usepackage[numbers,sort&compress]{natbib}

\usepackage[acronyms,nonumberlist,nopostdot,nomain,nogroupskip]{glossaries}
\usepackage{tablefootnote}
\usepackage{booktabs}

\usepackage{tabularx}

\usepackage{tikz}
\usepackage{pgfplots}
\pgfplotsset{compat=newest}
\pgfplotsset{plot coordinates/math parser=false}
\newlength\fheight
\newlength\fwidth
\usetikzlibrary{plotmarks,patterns,decorations.pathreplacing,backgrounds,calc,arrows,arrows.meta,spy,matrix}
\usepgfplotslibrary{patchplots,groupplots,colorbrewer}
\usepackage{tikzscale}
\usepackage{hyperref}

\newif\ifexttikz
\exttikzfalse

\ifexttikz
	\usetikzlibrary{external}
\fi

\title{Deep Learning at the Physical Layer: System Challenges and Applications to 5G and Beyond}
\author{\IEEEauthorblockN{
Francesco Restuccia, \emph{Member, IEEE}, and Tommaso Melodia, \emph{Fellow, IEEE}\vspace{-1cm}
\thanks{F. Restuccia and T. Melodia are with the Institute for the Wireless Internet of Things, Northeastern University, Boston, MA USA. Authors emails: \{frestuc, melodia\}@northeastern.edu. This paper has been accepted for publication in IEEE Communications Magazine.}
}\\
}

\usepackage{multirow}
\usepackage{tkz-kiviat}

\usepackage[font=footnotesize]{subcaption}
\usepackage[font=footnotesize]{caption}

\usepackage{mathtools}

\newacronym{6g}{6G}{sixth generation}
\newacronym{3gpp}{3GPP}{3rd Generation Partnership Project}
\newacronym{adc}{ADC}{Analog to Digital Converter}
\newacronym{5g}{5G}{5th generation}
\newacronym{aimd}{AIMD}{Additive Increase Multiplicative Decrease}
\newacronym{am}{AM}{Acknowledged Mode}
\newacronym{amc}{AMC}{Adaptive Modulation and Coding}
\newacronym{aoa}{AoA}{Angle of Arrival}
\newacronym{aod}{AoD}{Angle of Departure}
\newacronym{aqm}{AQM}{Active Queue Management}
\newacronym{awgn}{AGWN}{Additive White Gaussian Noise}
\newacronym{balia}{BALIA}{Balanced Link Adaptation}
\newacronym{bdp}{BDP}{Bandwidth-Delay Product}
\newacronym{bf}{BF}{Beamforming}
\newacronym{cc}{CC}{Congestion Control}
\newacronym{cdf}{CDF}{Cumulative Distribution Function}
\newacronym{cn}{CN}{Core Network}
\newacronym{cnn}{CNN}{convolutional neural network}
\newacronym{dnn}{DNN}{deep neural network}
\newacronym{cqi}{CQI}{Channel Quality Information}
\newacronym{cp}{CP}{Control Plane}
\newacronym{csirs}{CSI-RS}{Channel State Information - Reference Signal}
\newacronym{dc}{DC}{Dual Connectivity}
\newacronym{dce}{DCE}{Direct Code Execution}
\newacronym{drl}{DRL}{Deep Reinforcement Learning}
\newacronym{dci}{DCI}{Downlink Control Information}
\newacronym{dmr}{DMR}{Deadline Miss Ratio}
\newacronym{dmrs}{DMRS}{DeModulation Reference Signal}
\newacronym{e2e}{E2E}{End-to-End}
\newacronym{ecn}{ECN}{Explicit Congestion Notification}
\newacronym{ebs}{EBS}{exhaustive beam sweep}
\newacronym{edf}{EDF}{Earliest Deadline First}
\newacronym{enb}{eNB}{evolved Node Base}
\newacronym{epc}{EPC}{Evolved Packet Core}
\newacronym{es}{ES}{Edge Server}
\newacronym{fdma}{FDMA}{Frequency Division Multiple Access}
\newacronym{fdd}{FDD}{Frequency Division Duplexing}
\newacronym[firstplural=Radio Access Technologies (RATs)]{rat}{RAT}{Radio Access Technology}
\newacronym{fs}{FS}{Fast Switching}
\newacronym{5gb}{5GB}{5G and beyond}
\newacronym{txer}{TX}{transmitter}
\newacronym{rxer}{RX}{receiver}
\newacronym{bt}{BT}{beam tracking}
\newacronym{ftp}{FTP}{File Transfer Protocol}
\newacronym{gnb}{gNB}{Next Generation Node Base}
\newacronym{bs}{BS}{Base Station}
\newacronym{harq}{HARQ}{Hybrid Automatic Repeat reQuest}
\newacronym{hetnet}{HetNet}{Heterogeneous Network}
\newacronym{hh}{HH}{Hard Handover}
\newacronym{hol}{HOL}{Head-of-Line}
\newacronym{ia}{IA}{initial access}
\newacronym{imt}{IMT}{International Mobile Telecommunication}
\newacronym{phydl}{PHY-DL}{deep learning at the physical layer}
\newacronym{Phydl}{PHY-DL}{Physical-layer deep learning}
\newacronym{iot}{IoT}{Internet of Things}
\newacronym{los}{LOS}{Line-of-Sight}
\newacronym{lte}{LTE}{Long Term Evolution}
\newacronym{m2m}{M2M}{Machine to Machine}
\newacronym{ml}{ML}{machine learning}
\newacronym{dl}{DL}{deep learning}
\newacronym{mac}{MAC}{Medium Access Control}
\newacronym{mc}{MC}{Multi-Connectivity}
\newacronym{mcs}{MCS}{Modulation and Coding Scheme}
\newacronym{mec}{MEC}{Mobile Edge Cloud}
\newacronym{mi}{MI}{Mutual Information}
\newacronym{mimo}{MIMO}{Multiple Input, Multiple Output}
\newacronym{mmwave}{mmWave}{millimeter wave}
\newacronym{mmWave}{mmWave}{Millimeter wave}
\newacronym{mptcp}{MPTCP}{Multipath TCP}
\newacronym{mr}{MR}{Maximum Rate}
\newacronym{mss}{MSS}{Maximum Segment Size}
\newacronym{mtd}{MTD}{Machine-Type Device}
\newacronym{mtu}{MTU}{Maximum Transmission Unit}
\newacronym{nfv}{NFV}{Network Function Virtualization}
\newacronym{nlos}{NLOS}{Non-Line-of-Sight}
\newacronym{nr}{NR}{New Radio}
\newacronym{ofdm}{OFDM}{Orthogonal Frequency Division Multiplexing}
\newacronym{pdcch}{PDCCH}{Physical Downlink Control Channel}
\newacronym{pdcp}{PDCP}{Packet Data Convergence Protocol}
\newacronym{pdsch}{PDSCH}{Physical Downlink Shared Channel}
\newacronym{pdu}{PDU}{Packet Data Unit}
\newacronym{pf}{PF}{Proportional Fair}
\newacronym{pgw}{PGW}{Packet Gateway}
\newacronym{phy}{PHY}{physical layer}
\newacronym{pbch}{PBCH}{Physical Broadcast Channel}
\newacronym[plural=\gls{mme}s,firstplural=Mobility Management Entities (MMEs)]{mme}{MME}{Mobility Management Entity}
\newacronym{prb}{PRB}{Physical Resource Block}
\newacronym{pss}{PSS}{Primary Synchronization Signal}
\newacronym{pucch}{PUCCH}{Physical Uplink Control Channel}
\newacronym{pusch}{PUSCH}{Physical Uplink Shared Channel}
\newacronym{rach}{RACH}{Random Access Channel}
\newacronym{ran}{RAN}{Radio Access Network}
\newacronym{red}{RED}{Random Early Detection}
\newacronym{rf}{RF}{Radio Frequency}
\newacronym{rlc}{RLC}{Radio Link Control}
\newacronym{rlf}{RLF}{Radio Link Failure}
\newacronym{rrc}{RRC}{Radio Resource Control}
\newacronym{rrm}{RRM}{Radio Resource Management}
\newacronym{rr}{RR}{Round Robin}
\newacronym{rs}{RS}{Remote Server}
\newacronym{rsrp}{RSRP}{Reference Signal Received Power}
\newacronym{rss}{RSS}{Received Signal Strength}
\newacronym{rtt}{RTT}{Round Trip Time}
\newacronym{rw}{RW}{Receive Window}
\newacronym{rx}{RX}{Receiver}
\newacronym{sa}{SA}{standalone}
\newacronym{sack}{SACK}{Selective Acknowledgment}
\newacronym{sap}{SAP}{Service Access Point}
\newacronym{ap}{AP}{Access Point}
\newacronym{sch}{SCH}{Secondary Cell Handover}
\newacronym{scoot}{SCOOT}{Split Cycle Offset Optimization Technique}
\newacronym{sdma}{SDMA}{Spatial Division Multiple Access}
\newacronym{sinr}{SINR}{Signal to Interference plus Noise Ratio}
\newacronym{sm}{SM}{Saturation Mode}
\newacronym{snr}{SNR}{Signal-to-Noise-Ratio}
\newacronym{son}{SON}{Self-Organizing Network}
\newacronym{ss}{SS}{Synchronization Signal}
\newacronym{ssbs}{SSBs}{synchronization signal blocks}
\newacronym{ssb}{SSB}{synchronization signal block}
\newacronym{srs}{SRS}{Sounding Reference Signal}
\newacronym{sss}{SSS}{Secondary Synchronization Signal}
\newacronym{tb}{TB}{Transport Block}
\newacronym{tcp}{TCP}{Transmission Control Protocol}
\newacronym{tdd}{TDD}{Time Division Duplexing}
\newacronym{tdma}{TDMA}{Time Division Multiple Access}
\newacronym{tfl}{TfL}{Transport for London}
\newacronym{tm}{TM}{Transparent Mode}
\newacronym{trp}{TRP}{Transmitter Receiver Pair}
\newacronym{tti}{TTI}{Transmission Time Interval}
\newacronym{ttt}{TTT}{Time-to-Trigger}
\newacronym{tx}{TX}{Transmitter}
\newacronym{ue}{UE}{User Equipment}
\newacronym{ul}{UL}{Uplink}
\newacronym{uml}{UML}{Unified Modeling Language}
\newacronym{um}{UM}{Unacknowledged Mode}
\newacronym{utc}{UTC}{Urban Traffic Control}
\newacronym{vm}{VM}{Virtual Machine}
\newacronym{rsrq}{RSRQ}{Reference Signal Received Quality}
\newacronym{rssi}{RSSI}{Received Signal Strength Indicator}
\newacronym{crs}{CRS}{Cell Reference Signal}
\newacronym{nsa}{NSA}{Non Stand Alone}
\newacronym{mrdc}{MR-DC}{Multi \gls{rat} \gls{dc}}
\newacronym{endc}{EN-DC}{E-UTRAN-\gls{nr} \gls{dc}}
\newacronym{5gc}{5GC}{5G Core}
\newacronym{si}{SI}{Study Item}
\newacronym{iab}{IAB}{Integrated Access and Backhaul}
\newacronym{wf}{WF}{Wired-first}
\newacronym{hqf}{HQF}{Highest-quality-first}
\newacronym{pa}{PA}{Position-aware}
\newacronym{mlr}{MLR}{Maximum-local-rate}
\newacronym{wbf}{WBF}{Wired Bias Function}
\newacronym{mib}{MIB}{Master Information Block}
\newacronym{sib}{SIB}{Secondary Information Block}
\newacronym{kpi}{KPI}{Key Performance Indicator}
\newacronym{ppp}{PPP}{Poisson Point Process}
\newacronym{gtp}{GTP}{GPRS Tunneling Protocol}
\newacronym{amf}{AMF}{Access and Mobility Management Function}
\newacronym{dash}{DASH}{Dynamic Adaptive Streaming over HTTP}
\newacronym{http}{HTTP}{HyperText Transfer Protocol}
\newacronym{qos}{QoS}{Quality of Service}
\newacronym{udp}{UDP}{User Datagram Protocol}
\newacronym{cu}{CU}{Central Unit}
\newacronym{du}{DU}{Distributed Unit}
\newacronym{mt}{MT}{Mobile Termination}
\newacronym{sdap}{SDAP}{Service Data Adaptation Protocol}
\newacronym{tdm}{TDM}{Time Division Multiplexing}
\newacronym{fdm}{FDM}{Frequency Division Multiplexing}
\newacronym{sdm}{SDM}{Space Division Multiplexing}
\newacronym{dag}{DAG}{Directed Acyclic Graph}
\newacronym{st}{ST}{Spanning Tree}
\newacronym{ummimo}{UM-MIMO}{Ultra-massive Multiple Input, Multiple Output}
\newacronym{uavs}{UAVs}{Unmanned Aerial Vehicles}
\newacronym{wlan}{WLAN}{Wireless LAN}
\newacronym{rlnc}{RLNC}{Random Linear Network Coding}
\newacronym{drx}{DRX}{Discontinuous Reception}
\newacronym{cpu}{CPU}{Central Processing Unit}
\newacronym{txb}{TXB}{transmitter's beam}
\newacronym{rxb}{RXB}{receiver's beam}
\newacronym{dsp}{DSP}{digital signal processing}

\definecolor{desireRed}{RGB}{230,57,60}%
\definecolor{darkPurple}{RGB}{59,31,43}%
\definecolor{springGreen}{RGB}{37,223,145}%
\definecolor{queenBlue}{RGB}{69,123,157}%
\definecolor{spaceCadet}{RGB}{29,53,87}%

\makeglossaries

\begin{document}
\maketitle


\begin{abstract}
The unprecedented requirements of the \gls{iot}  have made fine-grained optimization of spectrum resources an urgent necessity. Thus, designing techniques able to extract knowledge from the spectrum in real time and select the optimal spectrum access strategy accordingly has become more important than ever. Moreover, \gls{5gb} networks will require complex management schemes to deal with problems such as adaptive beam management and rate selection. Although \gls{dl} has been successful in modeling complex phenomena, commercially-available wireless devices are still very far from actually adopting learning-based techniques to optimize their spectrum usage. In this paper, we first discuss the need for real-time \gls{dl} at the physical layer, and then summarize the current state of the art and existing limitations. We conclude the paper by discussing an agenda of research challenges and how \gls{dl} can be applied to address crucial problems in \gls{5gb} networks. \vspace{-0.7cm}
\end{abstract}

\glsresetall

\section{Introduction}

The wireless spectrum is undeniably one of nature's most complex phenomena. This is especially true in the highly-dynamic context of the \gls{iot}, where the widespread presence of tiny embedded wireless devices seamlessly connected to people and objects will make spectrum-related quantities such as fading, noise, interference, and traffic patterns hardly predictable with traditional mathematical models. Techniques able to perform real-time fine-grained spectrum optimization will thus become fundamental to squeeze out any spectrum resource available to wireless devices.

\begin{figure}[!h]
    \centering
    \includegraphics[width=0.9\linewidth]{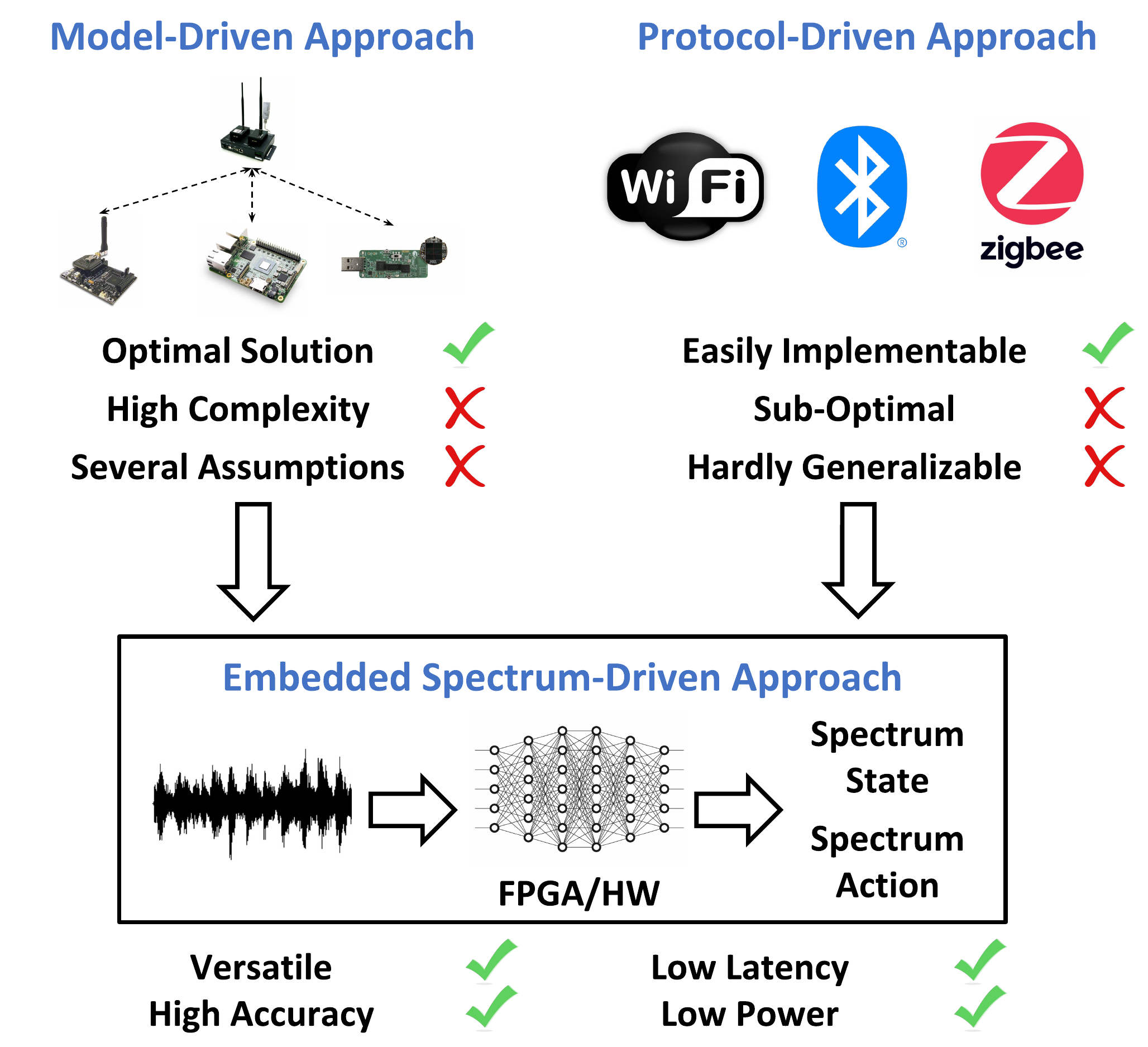}
    \caption{Key issues in today's wireless optimization approaches.}
    \label{fig:today}
    \vspace{-0.4cm}
\end{figure}

There are a number of key issues -- summarized in Fig.~\ref{fig:today} -- that make existing wireless optimization approaches not completely suitable to address the spectrum challenges mentioned above. On one hand, \emph{model-driven} approaches aim at (i) mathematically formalize the entirety of the network, and (ii) optimize an objective function. Although yielding optimal solutions, these approaches are usually NP-Hard, and thus, unable to be run in real time. Moreover, they rely on a series of modeling assumptions (\textit{e.g.}, fading/noise distribution, traffic and mobility patterns, and so on) that may not always be valid. On the other hand, \emph{protocol-driven} approaches consider a specific wireless technology (\textit{e.g.}, WiFi, Bluetooth or Zigbee) and attempt to heuristically change parameters such as modulation scheme, coding level, packet size, etc., based on metrics computed in real time from pilots and/or training symbols. Protocol-driven approaches, being heuristic in nature, necessarily yield sub-optimal performance.

To obtain the best of both worlds, a new approach called \emph{spectrum-driven} is being explored. In short, by using real-time \gls{ml} techniques implemented in the hardware portion of the wireless platform, we can design wireless systems that can \textit{learn by themselves} the optimal spectrum actions to take given  the current spectrum state. Concretely speaking, the big picture is to realize systems able to distinguish \emph{on their own} different spectrum states (\textit{e.g.}, based on noise, interference, channel occupation, and similar), and change their hardware and software fabric \textit{in real time} to implement the optimal spectrum action \cite{restuccia2020deepwierl,Restuccia-infocom2019}. However, despite the numerous recent advances, so far \textit{truly} self-adaptive and self-resilient cognitive radios have been elusive. On the other hand, the success of \gls{phydl} in addressing problems such as modulation recognition \cite{OShea-ieeejstsp2018}, radio fingerprinting \cite{restuccia2019deepradioid} and \gls{mac} \cite{Naparstek-ieeetwc2019} has taken us many steps in the right direction \cite{zhang2019deep}. Thanks to its unique advantages, \gls{dl} can really be a game-changer, especially when cast in the context of a real-time hardware-based implementation.


Existing work has mostly focused on generating spectrum data and training models in the cloud. However, a number of key \textit{system-level issues} still remain substantially unexplored. To this end, we notice that the most relevant survey work \cite{huang2019deep,o2017introduction} introduces research challenges from a learning perspective only. Moreover,   \cite{Mao-ieeecommsurtut2018} and similar survey work focuses on the application of DL to upper layers of the network stack. Since \gls{dl} was not conceived having the constraints and requirements of wireless communications in mind, it is still unclear what are the fundamental limitations of \gls{phydl}. Moreover, existing work has still not explored how \gls{phydl} can be used to address problems in the context of \gls{5gb} networks. For this reason, the first key contribution of this paper is to  discuss the research challenges of real-time \gls{phydl} without considering any particular frequency band or radio technology (Sections III and IV).  The second key contribution is the introduction of a series of practical problems that may be addressed by using \gls{phydl} techniques in the context of \gls{5gb} (Section IV-C). Notice that \gls{5gb} networks are expected to be heavily based on millimeter-wave (mmWave) and ultra-wideband communications, hence our focus on these issues. Since an exhaustive compendium of the existing work in \gls{phydl} is outside the scope of this manuscript, we refer the reader to \cite{Mao-ieeecommsurtut2018} for an excellent survey.


\vspace{-0.2cm}

\section{Why Deep Learning at the Physical Layer?}\label{sec:dl_phy}

\begin{figure}[!h]
    \centering
    \includegraphics[width=\linewidth]{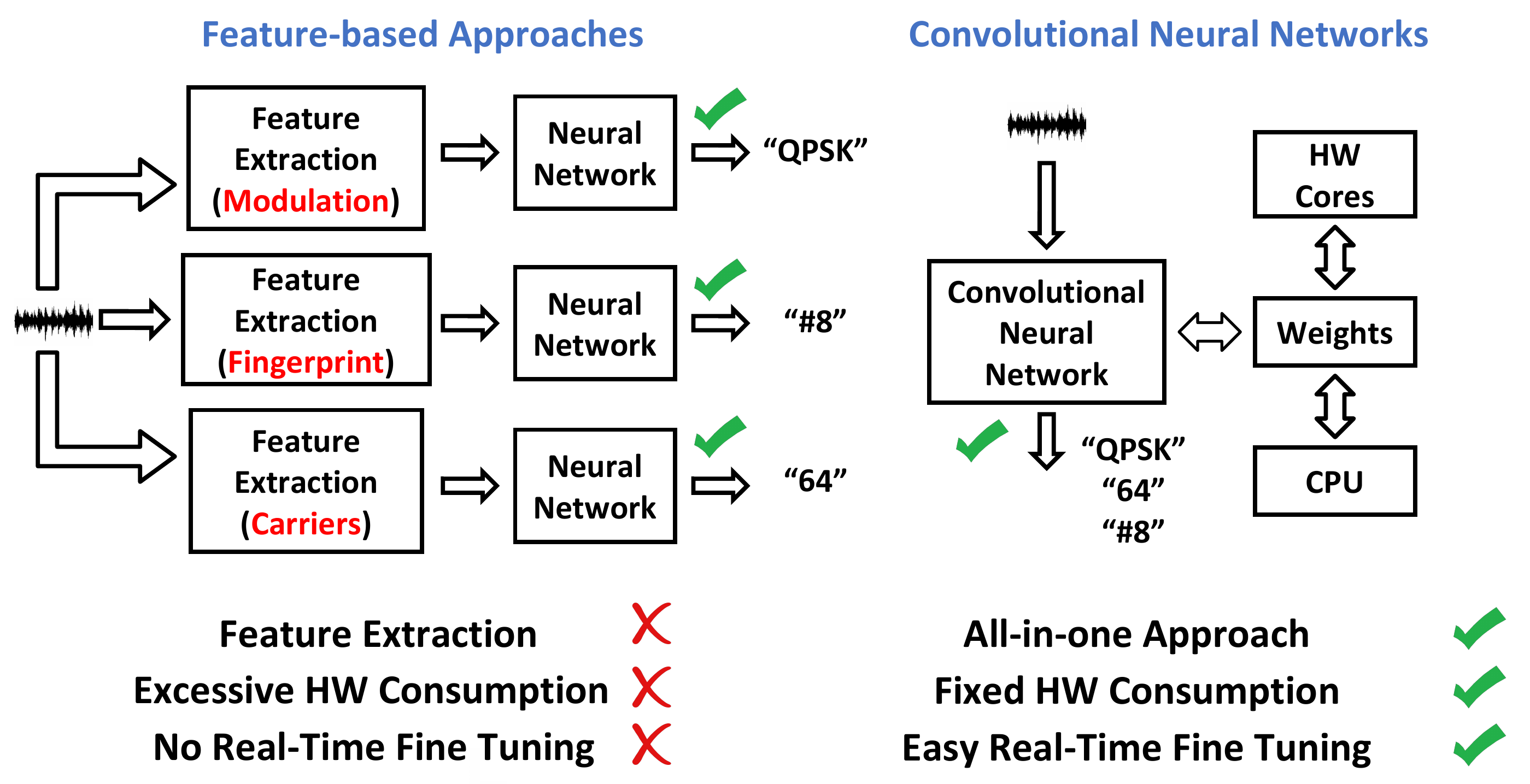}
    \caption{Feature-based Approaches vs Convolutional Neural Networks.}
    \label{fig:trad_vs_dl}
\end{figure}

\gls{dl} is exceptionally suited to address problems where closed-form mathematical expressions are difficult to obtain \cite{lecun2015deep}. For this reason, \glspl{cnn} are now being ``borrowed'' by wireless researchers to address handover and power management in cellular networks, dynamic spectrum access, resource allocation/slicing/caching, video streaming, and rate adaptation, just to name a few. Fig.~\ref{fig:trad_vs_dl} summarizes why traditional \gls{ml} may not effectively address real-time physical-layer problems. Overall, \gls{dl} relieves from the burden of finding the right ``features'' characterizing a given wireless phenomenon.~At the physical layer, this key advantage comes almost as a necessity for at least three reasons, which are discussed below. \vspace{0.1cm}

\emph{Highly-Dimensional Feature Spaces.}~Classifying waveforms ultimately boils down to distinguishing small-scale \textit{patterns} in the in-phase-quadrature (I/Q) plane, which may not be clearly separable in a low-dimensional feature space. For example, in radio fingerprinting we want to distinguish among hundreds (potentially thousands) of devices based on the unique imperfections imposed by the hardware circuitry. While legacy low-dimensional techniques can correctly distinguish up to a few dozens of devices \cite{12_vo2016fingerprinting}, \gls{dl}-based classifiers can scale up to hundreds of devices by learning extremely complex features in the I/Q space \cite{restuccia2019deepradioid}.~Similarly, O'Shea \emph{et al.} \cite{OShea-ieeejstsp2018} have demonstrated that on the 24-modulation dataset considered, \gls{dl} models achieve on the average about 20\% higher classification accuracy than legacy learning models under noisy channel conditions.

\emph{All-in-One Approach.}~The second key advantage of \gls{dl} is that automatic feature extraction allows the system designer to reuse the same \gls{dl} architecture -- and thus, the same hardware circuit -- to address different learning problems.  This is because, as we explain in Section III.B, \glspl{cnn} learn I/Q patterns in the I/Q plane, making them amenable to address different classification problems. Existing work, indeed, has demonstrated that CNNs can be used for very different problems, ranging from modulation recognition \cite{OShea-ieeejstsp2018} to radio fingerprinting \cite{restuccia2019deepradioid}. \glspl{cnn} also keeps latency and energy consumption constant, as explained in Section \ref{sec:dl_phy}. Fig.~\ref{fig:trad_vs_dl} shows an example where a learning system is trained to classify modulation, number of carriers and fingerprinting. While \gls{dl} can concurrently recognize the three parameters, traditional learning requires different feature extraction processes for each of the classification outputs. This, in turn, increases hardware consumption and hinders fine-tuning of the learning model.\smallskip 

\emph{Real-Time Fine Tuning.}~Model-driven optimization offers predictable performance only when the model actually matches the reality of the underlying phenomenon being captured. This implies that model-driven systems can yield sub-optimal performance when the model assumptions are different from what the network is actually experiencing. For example, a model assuming a Rayleigh fading channel can yield incorrect solutions when placed in a Rician fading environment. By using a \textit{data-driven} approach, \gls{phydl} may be easily fine-tuned through the usage of fresh spectrum data, which can be used to find in real time a better set of parameters through gradient descent. Hand-tailored model-driven systems may result to be hard to fine-tune, as they might depend on a set of parameters that are not easily adaptable in real time (e.g., channel model). While \gls{dl} ``easily'' accomplishes this goal by performing batch gradient descent on fresh input data, the same is not true for traditional \gls{ml}, where tuning can be extremely challenging since it would require  to completely change the circuit itself.\vspace{-0.3cm}

\section{Deep Learning at the Physical Layer:\\System Requirements and Challenges}\label{sec:requirements}

 The target of this section is to discuss existing system-level challenges in PHY-DL, as well as the state of the art in addressing these issues. For a more detailed compendium of the state of the art, the interested reader can take a look at the following comprehensive surveys~\cite{zhang2019deep,huang2019deep,o2017introduction,Mao-ieeecommsurtut2018}. \smallskip

\begin{figure}[!h]
    \centering
    \includegraphics[width=\linewidth]{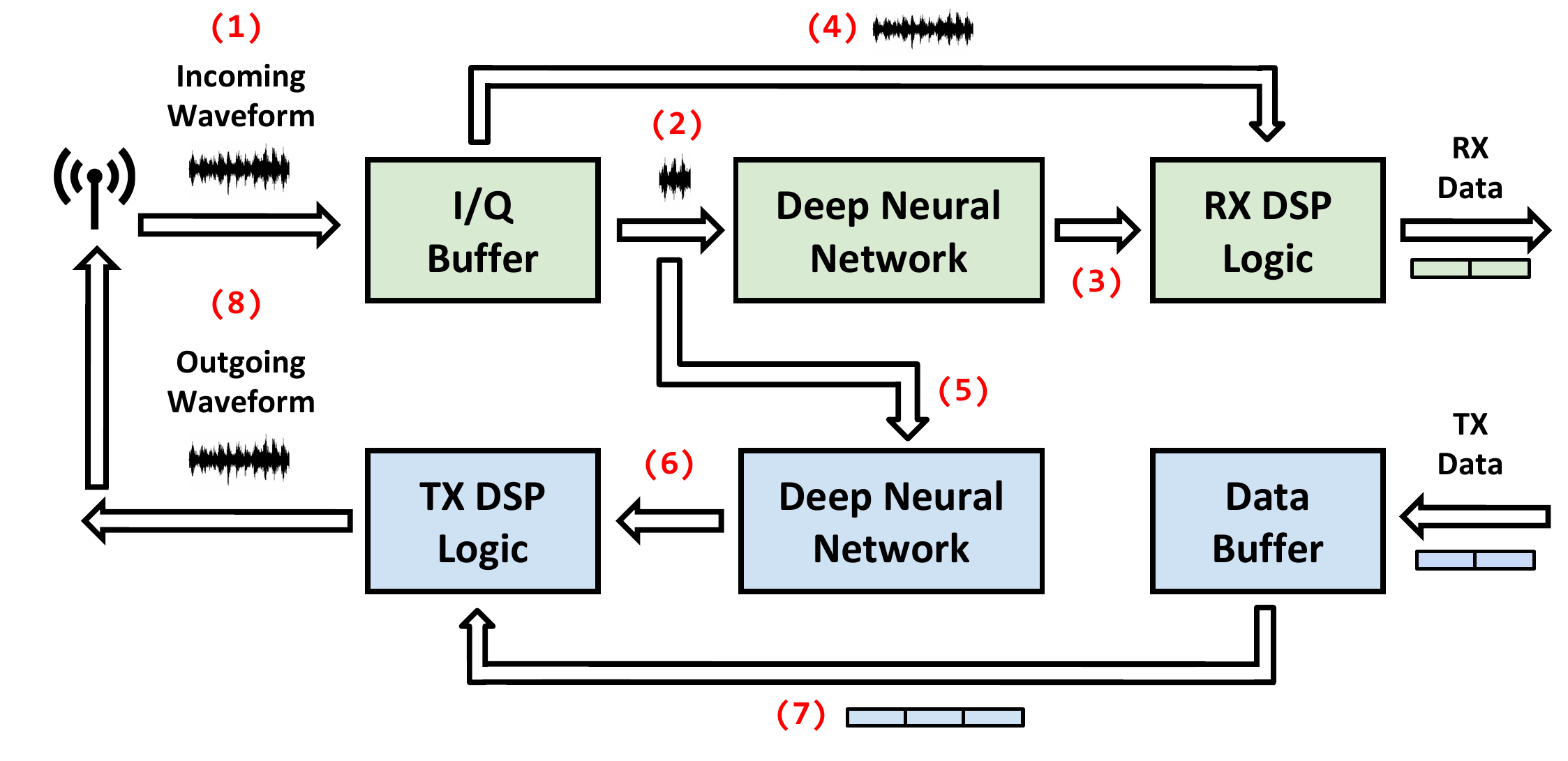}
    \caption{Main operations in a learning-based wireless device.}
    \label{fig:mainops}
\end{figure}

To ease the reader into the topic,  we summarize at a very high level the main components and operations of a learning-based  wireless device in Fig.~\ref{fig:mainops}. The core feature that distinguishes learning-based devices is that \gls{dsp} decisions are driven by \glspl{dnn}. In particular, in the \gls{rx} \gls{dsp} chain the incoming waveform is first received and placed in an I/Q buffer (step 1). Then, a portion of the I/Q samples are forwarded to the \gls{rx} \gls{dnn} (step 2), which produces an inference that is used to reconfigure the \gls{rx} \gls{dsp} logic (step 3). For example, if a QPSK modulation is detected instead of BPSK, the \gls{rx} demodulation strategy is reconfigured accordingly. Finally, the incoming waveform is released from the I/Q buffer and sent for demodulation (step 4). At the transmitter's side, the I/Q samples are sent to the RX DNN and to the \gls{tx} \gls{dnn} to infer the current spectrum state (\textit{e.g.}, spectrum-wide noise/interference levels). As soon as the inference is produced and the \gls{tx} \gls{dsp} logic is changed (step 6), the \gls{tx}'s buffered data is released (step 7), processed by the \gls{tx} \gls{dsp} logic and sent to the wireless interface (step 8).

We identify three core challenges in \gls{phydl}, which are discussed below.\vspace{-0.4cm}

\subsection{Addressing Latency and Space Constraints}

Domains such as computer vision usually do not have extreme requirements in terms of maximum latency or number of weights of a \gls{dl} model.  This is also true when \gls{ml} is applied to higher layers of the protocol stack. For example, when we are uploading a picture on a social network, we do not expect a face recognition algorithm that automatically ``tags" us and our friends to run under a given number of milliseconds. The same happens when running a routing protocol, where few microseconds do not necessarily impact on the protocol's performance.  Very different, however, is the case of \gls{phydl}, where \gls{dsp} constraints and hardware limitations have to be heeded -- in some cases, down to the clock cycle level.\smallskip

\subsubsection{\textbf{System Challenges}}First, the \gls{dnn} must run quickly enough to avoid overflowing the I/Q buffer and/or the data buffer (see Fig.~\ref{fig:mainops}). For example, an incoming waveform sampled at 40~$\mathrm{MHz}$ (\textit{e.g.}, a WiFi channel) will generate a data stream of 160$~\mathrm{MB/s}$, provided that each I/Q sample is stored in a 4-byte word. With an I/Q buffer of 1~kB, the \gls{dnn} must run with a latency less than 6.25~us to avoid buffer overflow. 

Moreover, the \gls{dnn} must be fast enough to be (much) less than the channel's coherence time and the transmitter's frequency in changing parameters. For example, if the channel coherence time is 10ms, the \gls{dnn} should run with latency \emph{much less} than 10ms to make meaningful inference. However, if the transmitter switches modulation every 1ms, the \gls{dnn} has to run with latency less than 1ms if it wants to detect modulation changes. The examples clearly show that lower \gls{dnn} latency implies (i) higher admissible sampling rate of the waveform, and thus, higher bandwidth of the incoming signal; (ii) higher capability of analyzing fast-varying channels. 

Hardware resource utilization is a spinous issue. Nowadays, \gls{dl} models usually have tens of millions of parameters, \textit{e.g.}, AlexNet has some 60M weights while VGG-16 about 138M. Obviously, it is hardly feasible to entirely fit these models into the hardware fabric of even the most powerful embedded devices currently available. Moreover, it is not feasible to run them from the cloud and transfer the result to the platform due to the additional delay involved. Therefore, \gls{phydl} have also to be relatively small to be feasibly implemented on embedded devices.  Resource utilization also directly impact energy consumption, which is a critical resource in embedded systems. Indeed, the more area (\textit{i.e.}, look-up tables, block RAMs, and so on) the model occupies in the hardware, the more the energy consumption.  However, it has been shown \cite{Restuccia-infocom2019} that thanks to the lower latency, implementing the model in a field-programmable gate array (FPGA) can lead to substantial energy reduction (up to 15x) with respect to an implementation in the central processing unit (CPU).\smallskip

\subsubsection{\textbf{Existing Work}}In \cite{Restuccia-infocom2019}, the authors propose \emph{RFLearn}, a hardware/software framework to integrate a Python-level \gls{cnn} into the \gls{dsp} chain of a radio receiver. The framework is based on high-level synthesis (HLS) and translates the software-based \gls{cnn} to an FPGA-ready circuit. Through HLS, the system constraints on accuracy, latency and power consumption can be tuned based on the application. As a practical case study, the authors train  several models to address the problem of modulation recognition, and show that latency and power consumption can be reduced by 17x and 15x with respect to a model running in the CPU. Moreover, it is shown that accuracy of over 90\% can be achieved with a model of only about 30,000 parameters. \gls{drl} techniques are integrated at the transmitter's side with \textit{DeepWiERL} \cite{restuccia2020deepwierl},  a hybrid software/hardware \gls{drl} framework to support the  training  and  real-time  execution  of  state-of-the-art  \gls{drl} algorithms on top of embedded devices. Moreover, DeepWiERL includes a novel supervised \gls{drl} model selection and bootstrap technique that  leverages HLS and transfer learning to orchestrate a \gls{dnn} architecture  that decreases convergence time and satisfies application and hardware constraints. \vspace{-0.4cm}

\subsection{Designing Features and Addressing Stochasticity}

In computer vision, \glspl{dnn} are trained to detect small-scale ``edges'' (\textit{i.e.}, contours of eyes, lips, etc.), which become more and more complex as the network gets deeper (\textit{i.e.}, mouth, eyes, hair type, etc.).  This is precisely the property that makes these networks excellent at detecting, \textit{e.g.}, an object or a face in an image, irrespective of where it occurs. In the wireless domain, CNNs do not operate on images but on I/Q samples, thus input tensors must be constructed out of I/Q samples. 

\begin{figure}[!h]
    \centering
    \includegraphics[width=0.95\linewidth]{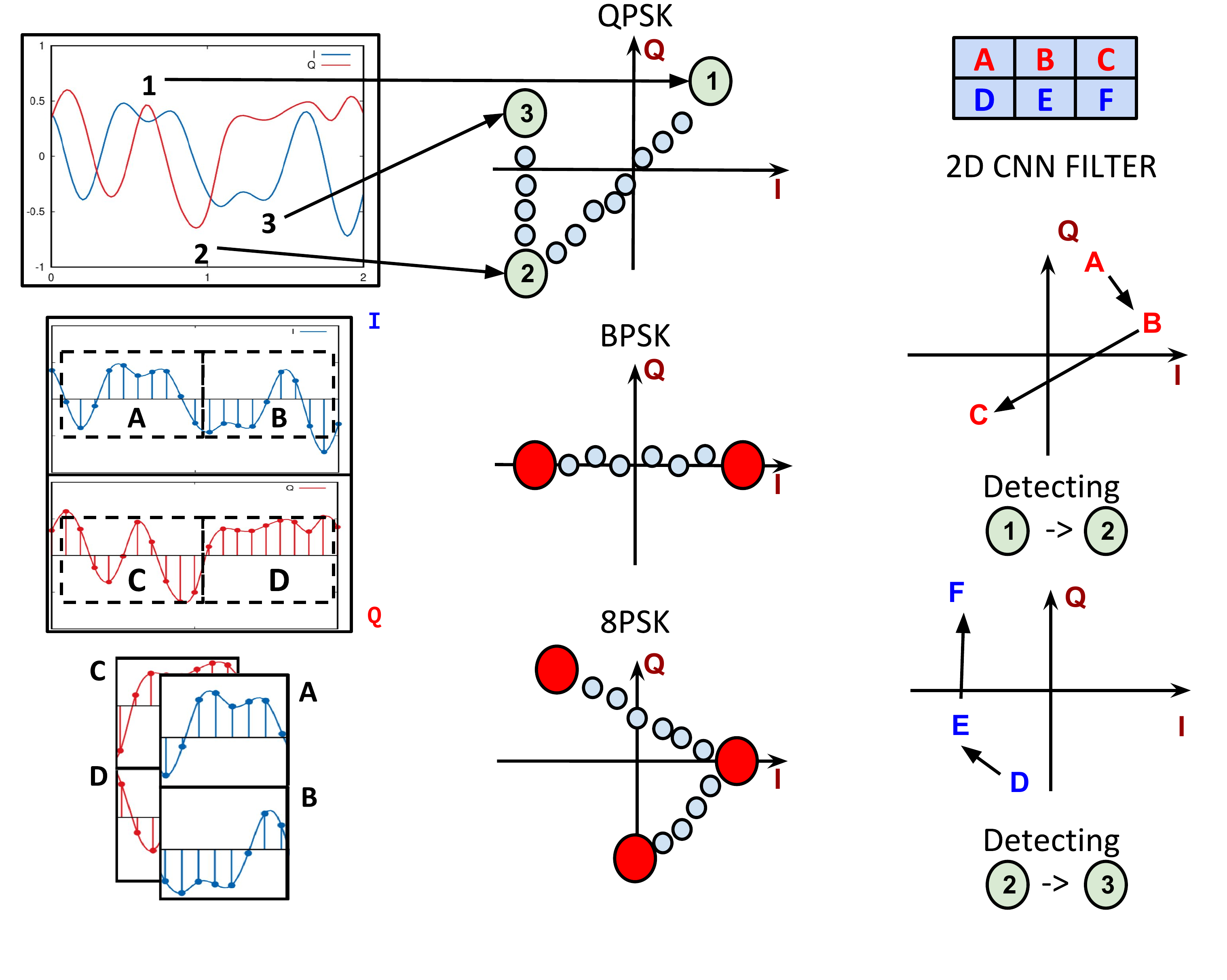}
    \caption{On the left, we show how to construct an input tensor of size (in this example, 10x10x2) from an I/Q waveform. In the center, we show various examples of how a waveform corresponds to transitions in the I/Q complex plane, for QPSK, BPSK and 8PSK modulations. On the right side, we show how a 2x3 filter of a CNN can detect to distinguish the transition between the first, second and third symbol of a modulation.\vspace{-0.4cm}}
    \label{fig:tensors}
\end{figure}

To make an example, the left side of Fig.~\ref{fig:tensors} shows the approach based on two-dimensional (2D) convolution proposed in \cite{Restuccia-infocom2019}. Specifically, input tensors are constructed by ``stacking up'' $H$ rows of $W$ consecutive I/Q samples. Fig.~\ref{fig:tensors} shows examples of transitions in the I/Q complex plane corresponding to QPSK, BPSK, and 8PSK. The transitions corresponding to the points (1) to (3) are shown in the upper-left side of Fig.~\ref{fig:tensors}. The figure clearly shows that different modulation waveforms present different I/Q transition patterns. For example, the transitions between (1, 0) and (-1, 0) peculiar to BPSK do not appear in QPSK, which presents a substantially different constellation.  This can constitute a unique ``signature'' of the signal that can  eventually be learned by the \gls{cnn} filters. The right side of Fig.~\ref{fig:tensors} shows an example of a 2x3 filter in the first layer of a \gls{cnn} trained for BPSK vs QPSK modulation recognition. Specifically, the first row of the filter (\textit{i.e.}, A, B, C) detects I/Q patterns where the waveform transitions from the first to the third quadrant (which correspond to the symbol ``1" to ``2" transition in our example) while the second row (\textit{i.e.}, D, E, F) detects transitions from the third to the second quadrant (which correspond to the symbol ``2" to ``3" transition).

\begin{figure}[!h]
    \centering
    \includegraphics[width=\linewidth]{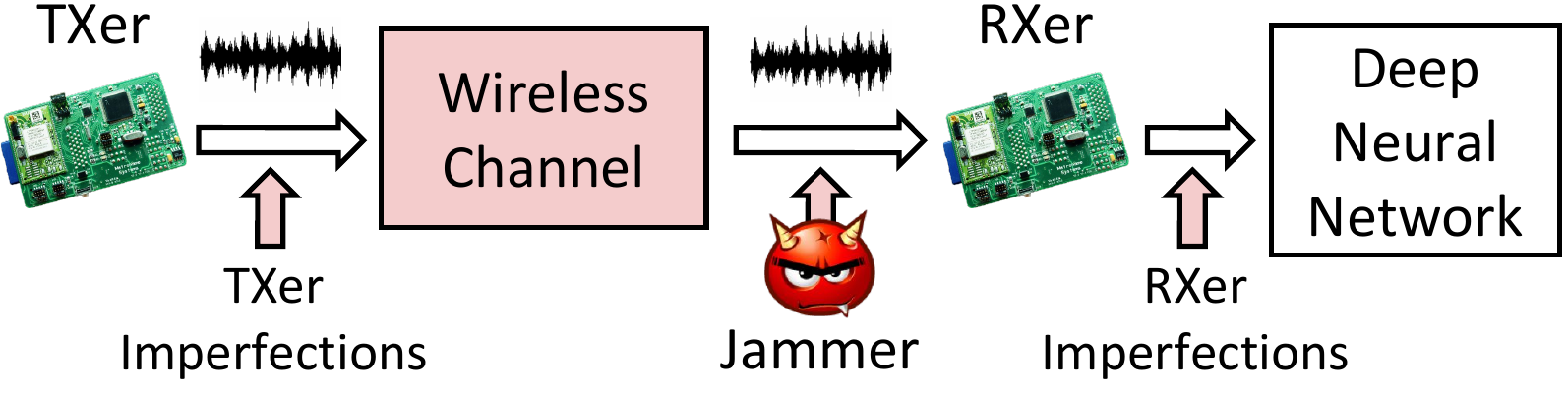}
    \caption{Source of randomness in \gls{phydl}.}
    \label{fig:source}
    \vspace{-0.2cm}
\end{figure}

However, the above and similar \gls{cnn}-based approaches \cite{OShea-ieeejstsp2018} do not fully take into account that a \gls{phydl} system is inherently \textit{stochastic} in nature. The first one is the unavoidable noise and fading that is inherent to any wireless transmission. Although channel statistics could be stationary in some cases, (i) these statistics cannot be valid in every possible network situation; (ii) a \gls{cnn} cannot be trained on all possible channel distributions and related realizations; (iii)  a \gls{cnn} is hardly re-trainable in real-time due to its sheer size. Recent research \cite{restuccia2019deepradioid} has shown that the wireless channel makes it highly unlikely to deploy \gls{dl} algorithms that will function without periodic fine-tuning of the weights \cite{ditzler2015learning}. Fig.~\ref{fig:source} summarizes the main sources of randomness in \gls{phydl}.

The second factor to consider is adversarial action (\textit{i.e.}, jamming), which may change the received signal significantly and usually, in a totally unpredictable way. The third factor is the unavoidable imperfections hidden inside the RF circuitry of off-the-shelf radios (\textit{i.e.}, I/Q imbalance, frequency/sampling offsets, and so on). This implies that signal features can (and probably will in most cases) change over time, in some cases in a very significant way. \smallskip

\subsubsection{\textbf{Existing Work}}

The issue of \gls{phydl} stochasticity has been mostly investigated in the context of radio fingerprinting \cite{shawabka2020exposing}. Specifically, the authors collected more than 7TB of wireless data obtained from 20 bit-similar wireless devices over the course of 10 days in different environments. The authors show that the wireless channel decreases the accuracy from 85\% to 9\%. However, another key insight is that waveform equalization can increase the accuracy by up to 23\%. To address the issue of stochasticity, the \emph{DeepRadioID} system \cite{restuccia2019deepradioid} was recently proposed, where finite input response filters (FIRs) are  computed at the receiver's side to compensate current channel conditions by being applied at the transmitter's side. The authors formulated a \textit{Waveform Optimization Problem} (WOP) to find the optimum FIR for a given CNN. Since the FIR is tailored to the specific device's hardware, it is shown that an adversary is not able to use a stolen FIR to imitate a legitimate device's fingerprint. The \emph{DeepRadioID} system was evaluated  with a testbed of 20 bit-similar software-defined radios (SDRs), as well as on two datasets containing transmissions from 500 ADS-B devices and by 500 WiFi devices. Experimental results show that \emph{DeepRadioID} improves the fingerprinting accuracy by 27\% with respect to the state of the art.\vspace{-0.2cm}

\begin{figure*}
    \centering
    \includegraphics[width=0.95\linewidth]{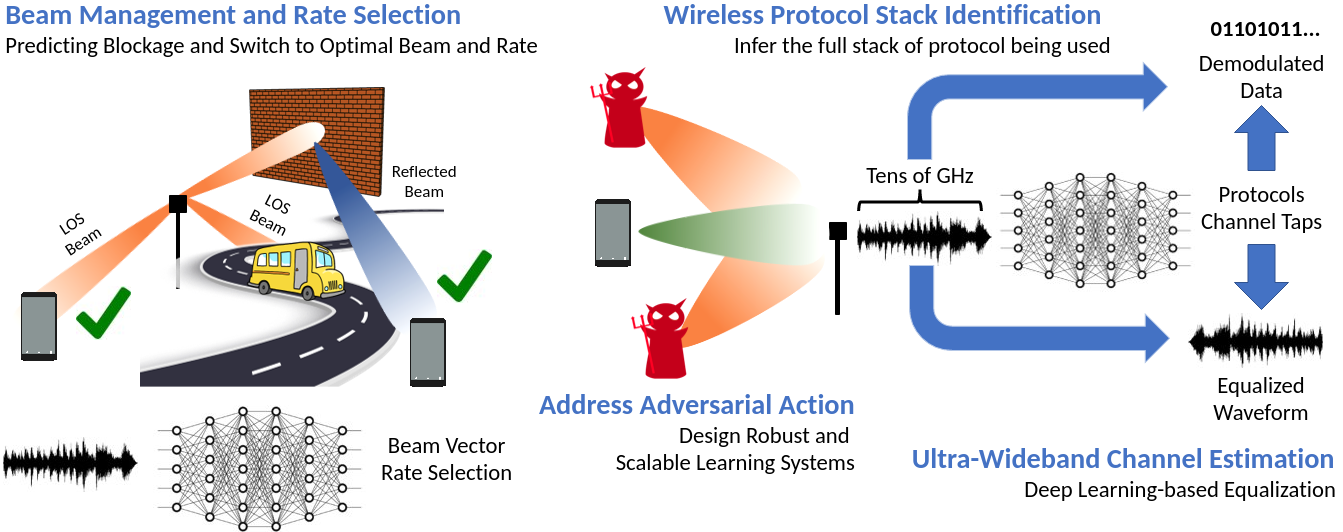}
    \caption{Summary of Main Research Challenges in \gls{phydl} and Applications to \gls{5gb} Networks.}
    \label{fig:summary}
    \vspace{-0.2cm}
\end{figure*}


\section{Deep Learning at the Physical-Layer:\\The Way Ahead}\label{sec:way_ahead}

We now present an agenda of research opportunities in \gls{phydl}. Fig.~\ref{fig:summary} summarizes the challenges discussed below. \vspace{-0.4cm}

\subsection{ Large-scale Experiments and Data Collection}

So far, \gls{phydl} techniques have been validated in controlled, lab-scale environments and with a limited number of wireless technologies. Although large-scale datasets in the area or radio fingerprinting have been produced, other \gls{phydl} problems (\textit{e.g.}, modulation recognition) have been clearly left behind. For this reason, the research community desperately needs large-scale experimentation to really understand whether these techniques can be applied in realistic wireless ecosystems where hundreds of nodes, protocols and channels will necessarily coexist.  Moreover, due to the current lack of common datasets, today every paper in the wireless \gls{ml} domain can claim to be ``better than the previous one" in terms of accuracy.  For this reason, the creation of large-scale datasets shared with the research community at large should also be considered as a priority.

To bring \gls{phydl} to the next level, we need ``wireless data factories" able to generate I/Q data at unseen scale.  The newly-developed \textit{Platforms for Advanced Wireless Research} (PAWR) will be fundamental in addressing the above challenge (\url{https://advancedwireless.org}). The PAWR program will develop four platforms  to be shared among the wireless research community. The platforms will enable sub-6, millimeter-wave and drone experimentation capabilities in a multitude of real-world scenarios. Alongside PAWR, the Colosseum network emulator  (\url{http://experiments.colosseum.net}) will be soon open to the research community and  provide us with unprecedented data collection opportunities. Originally developed to support DARPA's spectrum collaboration challenge in 2019, Colosseum can emulate up to 256x256 4-tap wireless channels among 128 software-defined radios. Users can create their own wireless scenarios and thus create ``virtual worlds'' where learning algorithms can be truly stressed to their full capacity.\vspace{-0.4cm}

\subsection{Addressing Wireless Adversarial Learning} 

Up until now, researchers have focused on improving the accuracy of the \gls{phydl} model, without heeding security concerns. However, we know the accuracy of a \gls{dl} model can be significantly compromised by crafting adversarial inputs. The first kind of attack is called \textit{targeted}, where given a valid input, a classifier and a target class, it is possible to find an input close to the valid one such that the classifier is ``steered'' toward the target class. More recently, researchers have demonstrated the existence of \textit{universal perturbation vectors}, such that when applied to the majority of inputs, the classifier steers to a class different than the original one. On the other hand, the time-varying nature of the channel could compromise adversarial attempts. Moreover, the received waveforms still need to be decodable and thus cannot be extensively modified. Therefore, additional research is needed to fill the gap between AML and the wireless domain and  demonstrate \emph{if}, \emph{when}, and \emph{how} adversarial machine learning (AML) is concretely possible in practical wireless scenarios.\vspace{-0.1cm}

\subsection{Applications to 5G and Beyond}

 Below, we discuss a series of applications of \gls{dl} at the physical-layer to \gls{5gb}, and provide a roadmap of possible research avenues in the field.\smallskip

\textbf{Analyzing Ultra-wide Spectrum Bands.}~The millimeter wave (mmWave) and Terahertz (THz) spectrum bands have become the \emph{de facto} candidates for \gls{5gb}  communications. To fully unleash the power of these bands, mmwave/THz systems will operate with \textit{ultra-wide spectrum bands} -- in the order of several, perhaps tens of gigahertz (GHz). Thus, pilot-based channel estimation could not result to be the best strategy. Indeed, frequently transmitting pilots for the whole bandwidth can lead to severe loss of throughput. A neural network could be trained to infer the channel directly based on the I/Q samples, without requiring additional pilots. One possible strategy could be to leverage the packet headers or trailers as source of reference I/Q date to train the learning model. \smallskip

\textbf{Protocol Stack Identification.}~Next-generation networks will necessarily require fast and fine-grained optimization of parameters at all the layers of the protocol stack. Radios will thus need to be extremely spectrum-agile, meaning that wireless protocols should be used \textit{interchangeably} and according to the current spectrum circumstances.  To demodulate incoming waveforms transmitted with different strategies, it becomes necessary to \textit{infer} the waveform type -- and thus, the wireless protocol stack being used -- before feeding it to the \gls{dsp} logic.  To the best of our knowledge, this problem still remains open. Additional research should shed light on whether physical-layer I/Q samples can be used to infer the whole stack of a wireless protocol. One possible avenue could be to extend the input size of the model and learn more complex features. However, this could increase latency to unacceptable levels. An alternative could be to utilize an ensemble model where smaller submodels are trained to analyze different portions of the waveform. This would ultimately help the model generalize yet remaining under acceptable latency levels. \smallskip

\textbf{ Blockage Prediction and Beam Alignment.}~Another major challenge of mmWave and THz communications is the severe path and absorption loss (\textit{e.g.}, oxygen at 60 GHz). Moreover, mmWave and THz carriers cannot penetrate physical obstacles such as dust, rain, snow, and other opaque objects (people, building, transportation vehicles), making them highly susceptible to blockage. This key aspect will require high directionality of antenna radiations (\textit{i.e.}, beamforming), which will increase the transmission range but also introduce the compelling need for proactive beam steering and rate adaptation techniques. Deep learning could be utilized to design  prediction techniques that can infer in real-time an incoming blockage in a beam direction and thus proactively ``steer'' the beam toward another direction.  In this spirit, Alrabeiah and Alkhateeb \cite{alrabeiah2020deep} have recently proven that under some conditions, we can leverage sub-6 GHz channels to predict the optimal mmWave beam and blockage status. Then, the authors develop a \gls{dl} model and test it using a publicly available dataset called DeepMIMO. However, DeepMIMO is obtained through simulations based on a ray tracer, and sub-6 channels may not be always available. Therefore, further research is needed to validate whether these approaches can be generalized to different channel conditions and obstacles.

 Regardless of obstacles, \gls{txer} and \gls{rxer} beams have to be perfectly aligned to maximize the \gls{snr} during the transmission. Usually, through pilot sequences, the \gls{rxer} is then able to compute the \gls{snr} for each of the possible TX-RX beam combinations. The complexity of these beam alignment techniques is thus quadratic in the number of beams. A possible approach we are currently exploring is \gls{phydl} of \emph{ongoing transmissions} between the \gls{txer} and other receivers to infer the current \gls{txer}'s  beam and thus align the \gls{rxer}'s beam with the \gls{txer}'s to avoid explicit beam scanning. We obtained some preliminary results  through our \gls{mmwave} testbed, where we train a \gls{cnn} to identify the \gls{txer}'s beam. We experimented with two 24-element phased array antennas, and with a 12-beam and 24-beam codebook. The results indicate that we are able to achieve accuracy close to 90\% in both cases, with a \gls{cnn} constituted by 7 convolutional layers (each with 64 kernels of size 1x7) and 2 dense layers of 128 neurons, with a total of 848,472 parameters.\smallskip


 \textbf{PHY Virtualization and Optimization.}~To deliver the required services, \gls{5gb} will strongly depend on \emph{virtualization} techniques, where PHY resources such as spectrum, transmission time, base stations, etc., networks will become shared among different virtual network operators (VNOs). This will allow seamless delivery of stringent Quality of Experience (QoE) requirements, such as real-time surveillance, web browsing, and high-quality video content delivery, among others. However, as the network size increases, the relationship between computing, storage and radio resources will be hard to model in explicit mathematical terms. To establish the resources, \gls{drl} could be utilized to learn representations of the current state of the system and tie them to optimal actions. Recently, the research community has started to move in this direction. Ayala-Romero \emph{et al.} \cite{Romero-vrain} presented a system where an autoencoder is used to project high-dimensional context data (such as traffic and signal quality patterns) into a lower-dimensional representation. Then, an actor-critic neural network structure is used to map (encoded) contexts into resource control decisions. However, the proposed system is single-agent only, and the only physical-layer decision is related to the modulation and coding scheme used. It is unclear yet whether \gls{drl} can be extended to more complex problems and to multi-agents in realistic scenarios. \vspace{-0.2cm}

\section{Conclusions} \label{sec:conclusions}

The unprecedented scale and complexity of today's wireless systems will necessarily require protocols and architectures to rely on data-driven techniques. In this paper, we have provided an overview of PHY-DL and the state of the art in this topic. We have also introduced a roadmap of exciting research opportunities, which are definitely not easy to tackle but that if addressed, will take PHY-DL to the next level. We hope that this paper will inspire and spur significant wireless research efforts in this  exciting field in the years to come.


\footnotesize
\bibliographystyle{IEEEtran}
\bibliography{bibl}

\begin{IEEEbiographynophoto}
{Francesco Restuccia} (M'16) is an Assistant Professor at Northeastern University, USA. His research interests lie in wireless systems and artificial intelligence. Dr. Restuccia has published extensively in these areas and serves as TPC member and reviewer for top IEEE and ACM venues. 
  
 \vspace{0.1cm}
 \noindent
 \textbf{Tommaso Melodia} (F'18) is the William Lincoln Smith Chair Professor at Northeastern University. He is the Director of Research for the PAWR Project Office. He is the Editor in Chief of Computer Networks, an Associate Editor for the IEEE Transactions on Mobile Computing, and the IEEE Transactions on Biological, Molecular, and Multi-Scale Communications.
\end{IEEEbiographynophoto}

\end{document}